\documentclass[runningheads]{llncs}
\usepackage[T1]{fontenc}
\usepackage{graphicx}

\usepackage{soul}
\usepackage{tikz}
\newcommand{\tikzxmark}{
\tikz[scale=0.23] {
    \draw[line width=0.7,line cap=round] (0,0) to [bend left=6] (1,1);
    \draw[line width=0.7,line cap=round] (0.2,0.95) to [bend right=3] (0.8,0.05);
}}
\newcommand{\tikzcmark}{
\tikz[scale=0.23] {
    \draw[line width=0.7,line cap=round] (0.25,0) to [bend left=10] (1,1);
    \draw[line width=0.8,line cap=round] (0,0.35) to [bend right=1] (0.23,0);
}}

\usepackage{xcolor}
\usepackage{amsmath, systeme}
\usepackage{url}
\usepackage{booktabs}
\usepackage{array}
\usepackage{multirow}
\usepackage{xspace}

\newlength{\colA}
\newlength{\colB}
\newlength{\colC}
\newlength{\colD}
\newlength{\colE}

\newcommand{\WideCellThree}[1]{\multicolumn{3}{p{\dimexpr\colC+\colD+\colE+4\tabcolsep\relax}}{#1}}

\graphicspath{{figures}}
\newcommand{\crow}{Crow\texttrademark{}\xspace}

\newcommand{\im}{IronMiner\xspace}
\newcommand{\cvss}[1]{\ensuremath{\text{CVSS}_{v#1}}}
\newcommand{\yepss}{\ensuremath{\text{EPSS}_{365}}\xspace}

\begin{document}

\title{When Cyber Scoring Systems Diverge: An Empirical Comparison}
\titlerunning{When Cyber Scoring Systems Diverge}

\author{
Kirsi Hellsten\inst{1} \and
Joni Herttuainen\inst{1} \and
Ambrose Kam\inst{2} \and
Arlanda Johnson\inst{2} \and
David Welsh\inst{2} \and
Kimmo K. Kaski\inst{1}
}

\authorrunning{K. Hellsten et al.}

\institute{
Aalto University School of Science,
P.O. Box 11000, 00076 Aalto, Finland \email{firstname.lastname@aalto.fi} \and
Lockheed Martin
\email{\{ambrose.kam,arlanda.e.johnson.jr,david.welsh\}@lmco.com}}

\maketitle

\begin{abstract}

Vulnerability scoring systems underpin cyber patch prioritization and risk management, but their comparative behavior is almost always assessed in the abstract, through correlation studies in IT vulnerability databases, rather than by the operational consequences they produce when embedded in a system-level risk model. Here we present an empirical comparison of four vulnerability scoring systems, namely CVSS (Common Vulnerability Scoring System), EPSS (Exploit Prediction Scoring System), SSVC (Stakeholder-Specific-Vulnerability Categorization), and \im (operationally calibrated proprietary scoring system). As a substrate for comparison, we use a reconstruction of the 2015 Ukraine Power Grid operational-technology (OT) network that provides a documented incident topology. The results show a high degree of disagreement between the scoring systems. This suggests that the choice of the scoring system could significantly influence mitigation strategies and vulnerability prioritization, implying that a composite or hybrid scoring approach could offer a more suitable solution.

\end{abstract}

\section{Introduction}

Vulnerability scoring is central in modern security operations of organizational network systems. With far more vulnerabilities disclosed for attackers to exploit than any organization can remediate, defenders rely on scores to decide what to patch first and on scanners, vulnerability databases, and risk-management programs built around them. The dominant standard, the Common Vulnerability Scoring System (CVSS), has gone through successive revisions, v2, v3.0, v3.1 and emerging v4, each refining how exploitability and impact are quantified. In addition to CVSS, several complementary and competing systems have become available. For example, the Exploit Prediction Scoring System (EPSS) estimates near-term exploitation likelihood from observed activity, Stakeholder-Specific Vulnerability Categorization (SSVC)  frames prioritization as a decision process, and a range of vendor and operationally calibrated systems including Lockheed Martin's \im that incorporates the intelligence of an attack incident, not captured by the CVSS metrics.

So far, comparisons between different scoring systems have been conducted in the abstract. The literature is dominated by correlation studies that align scores across large IT vulnerability databases and report aggregate agreement or disagreement. However, these analyzes reveal little about what matters to a defender, that is, the operational consequence of a scoring choice once it is embedded in a model of how an attack propagates through a specific network. They are also overwhelmingly IT-centric, while operational-technology (OT) and industrial control system (ICS) networks have distinctive topologies, long-lived legacy software, and exploitation dynamics in which lateral movement through bridging nodes, rather than the severity of any single vulnerability, often determines the impact of an attack. Whether these scoring systems agree or disagree in that setting and whether any disagreement is structured or merely noise has not yet been established empirically.

We address this gap using a reconstruction of the 2015 Ukraine Power Grid OT network as a documented test environment as a substrate for vulnerability scoring comparison. We performed
comparison for CVSS, EPSS, SSVC, and \im vulnerability scoring systems 
to study whether different scoring systems produce equivalent network-wide risk rankings, is the possible divergence systematic or random, do such divergences change defensive prioritization, and how to utilize this information.

The contributions of this paper are threefold. First, we provide empirical comparison of general-purpose and operationally calibrated scoring systems on a documented OT incident topology rather than an aggregate IT vulnerability corpus. Second, we quantify the disagreement between these systems, including between two versions of CVSS, at the level of individual CVEs, subscores, and network placement. Third, we present and discuss two hybrid formulations (a weighted composite score and a CVSS-anchored priority score) that operationalize the finding that these systems answer fundamentally different questions.

The remainder of the paper is organized as follows. Section 2 reviews related work. Section 3 presents the methods and materials, including the scoring systems, the Ukraine Power Grid use case, the study methodology, and the dataset. Section 4 presents the results, and Section 5 draws conclusions.

\section{Related Works}

Comparisons between vulnerability scoring systems have so far been dominated by correlation studies over large IT vulnerability databases. Tan et al. (\cite{Tan_2025}) analyzed roughly 132,000 CVSS-scored vulnerabilities and 270,000 EPSS records and found no clear correlation between severity and exploitation likelihood: a vulnerability can be highly severe yet unlikely to be exploited, or only moderately severe yet actively targeted. Tracking EPSS over a 360-day window, they further observed that scores are highly dynamic, on average 677 vulnerabilities changed their EPSS score per day, increases were more than three times as likely as decreases, and medium-severity vulnerabilities showed the highest proportion of increases, suggesting that vulnerabilities initially dismissed as "medium" may become increasingly attractive targets over time. They concluded that CVSS and EPSS provide complementary but uncorrelated signals, and that neither should be used alone for prioritization.

Koscinski et al. (\cite{Koscinski_2025}) extended this line of work to four systems (CVSS, EPSS, SSVC, and Microsoft's Exploitability Index) using 600 vulnerabilities from the April–July 2024 Patch Tuesday releases. Across several agreement measures (Spearman, Kendall, Cohen's Kappa, Krippendorff's Alpha) they found consistently weak agreement: only 5 vulnerabilities appeared in the top-100 lists of all four systems. The systems also failed in opposite directions. CVSS, SSVC, and the Exploitability Index placed too many vulnerabilities in high-priority categories, while EPSS flagged very few. EPSS’s predictive value also proved limited when validated against the Known Exploited Vulnerabilities (KEV) catalog: only 19.9\% of later-exploited vulnerabilities ever exceeded an EPSS score of 0.5, and more than 22\% had no EPSS score before exploitation. Disagreement persisted even within common weakness classes (e.g., CWE-122, CWE-416, CWE-125), leading the authors to argue that the systems are answering fundamentally different questions and that vulnerability management should move toward context-aware, multi-signal decision models.

Two further studies point in the same direction. Pinaev (\cite{Pinaev26}) proposes CVPS, a composite prioritization model combining CVSS, EPSS, KEV evidence, and business criticality, arguing that risk-based composites outperform CVSS-only prioritization, though the model awaits large-scale empirical validation. Wunder et al. (\cite{Wunder24}), surveying 196 professional CVSS users, showed that even within a single scoring system consistency is not achieved in practice: ratings diverged both between evaluators and within the same evaluator over time, largely due to ambiguous metric definitions such as Scope.

Taken together, prior work establishes that scoring systems disagree, but it does so in the abstract, through aggregate correlation and agreement statistics over IT-centric vulnerability databases. What remains unexamined is the operational consequence of that disagreement once a scoring system is embedded in a model of attack propagation through a specific network, particularly in OT environments where topology and lateral movement dominate outcomes. This is the gap the present study addresses.

\section{Methods and Materials}
\subsection*{Vulnerability Scoring Systems}
To calculate the overall vulnerability of an OT or ICS network with distinct topology and software components,
the likelihood of exploitation of each component's vulnerability, i.e. exploitability, needs to be established, the values of which we derived using four different scoring systems.
One of them is CVSS (Common Vulnerability Scoring System), an open standard that describes the severity of software vulnerabilities in a consistent and quantitative way (\cite{CVSS_2.0}, \cite{CVSS_3.0}, \cite{CVSS_3.1}). It aims to help organizations prioritize remediation by producing a severity rating (\emph{Low}, \emph{Medium}, \emph{High}, \emph{Critical}) and a score between 0.0 and 10.0.  It is widely used in vulnerability databases, security scanners, and risk management programs. All different CVSS versions are built around three metric groups, i.e., the base metrics, temporal metrics, and environmental metrics. The base metrics are intrinsic characteristics of the vulnerability that do not change over time, while the temporal metrics are based on factors, such as exploit availability and patches, that change over time. The environmental metrics describe the organization-specific impact, indicating the importance of the affected systems.

CVSS v2 was released in 2007 (\cite{CVSS_2.0}), and its advantage is that the base score is simple and
easy to calculate, thus it is still widely used but less precise for modern attack scenarios. The base metrics can be divided into two categories: exploitability metrics and impact metrics.  Exploitability describes how difficult it is to exploit a vulnerability if the attacker knows that it exists. The three metrics to calculate exploitability are the access vector (AV), access complexity (AC), and authentication (Au). The access vector reflects how the vulnerability is exploited and the possible values are \emph{Local}, \emph{Adjacent Network}, and \emph{Network}. Access complexity measures the complexity of an attack required to exploit the vulnerability once an attacker has gained access to the target system, with possible values being \emph{High}, \emph{Medium}, and \emph{Low}. The authentication metric measures the number of times an attacker must authenticate to a target to exploit a vulnerability, and the possible values are \emph{Multiple}, \emph{Single}, and \emph{None}. The formula for calculating
exploitability is as follows
\begin{equation*}
    Exploitability_{v2} = 20\cdot AV\cdot AC\cdot Au.
\end{equation*}
Impact metrics consist of confidentiality impact (\emph{C}), integrity impact (\emph{I}), and availability impact (\emph{A}) (\cite{CVSS_2.0}) and it measures the damage when a vulnerability is exploited. The confidentiality impact metric measures the impact on confidentiality of a successfully exploited vulnerability. The integrity impact measures the impact on the integrity of a successfully exploited vulnerability. The availability impact measures the impact on the availability of a successfully exploited vulnerability. All impact metrics have three possible values: \emph{None}, \emph{Partial}, and \emph{Complete}. The formula for the \emph{Impact} reads as follows
\begin{equation*}
    Impact_{v2} = 10.41\cdot (1-(1-C)(1-I)(1-A)).
\end{equation*}
The base score is the CVSS headline number that combines exploitability and impact (\cite{CVSS_2.0}). It also acts as an input for the temporal score function. The formula for the base score is
\begin{equation*}
Base_{v2} = \operatorname{round}\big((0.6 \cdot {Impact} + 0.4 \cdot {Exploitability} - 1.5) \cdot f({Impact})\big),
\end{equation*}
where $f(Impact)=0$ or $=1.176$ for $Impact=0$ or $>0$,
respectively.
\cvss{2} has been criticized for
the authentication metric being
considered vague and outdated, and the impact score lacking nuances. It suffers from network-centric bias, meaning that it is less accurate for client-side and cloud attacks. Furthermore, there is no clear distinction between user interaction and attacker interaction.

CVSS v3.0 was released in 2015 as
a major redesign to address the shortcomings of the previous version (\cite{CVSS_3.0}). The main improvements include a more accurate representation of modern attack techniques, better impact modeling, and clear separation between attacker effort and user involvement. Furthermore, it has more realistic severity scoring, physical attacks are properly modeled, an explicit user interaction metric (\emph{UI}) was added, and the authentication metric was replaced with the privileges required metric (\emph{PR}). In addition to the user interaction and privileges required metrics, new exploitability metrics were introduced: attack vector (\emph{av}), attack complexity (\emph{ac}), and \emph{Scope}. The attack vector metric reflects the context by which vulnerability exploitation is possible with the values \emph{Network}, \emph{Adjacent}, \emph{Local}, or \emph{Physical} (\cite{CVSS_3.0}). The attack complexity describes the conditions beyond the attacker's control that must exist to exploit the vulnerability, and it can have values \emph{High} or \emph{Low}. The privileges required metric describes the level of privileges an attacker must possess before successfully exploiting the vulnerability, with values being \emph{High}, \emph{Low}, or \emph{None}. The user interaction metric captures the requirement for a user to participate in the successful compromise of the vulnerable component, and its value is either \emph{Required} or \emph{None}. The new key concept, \emph{Scope}, indicates whether a vulnerability can affect resources beyond its security authority, and its possible values are \emph{Changed} or \emph{Unchanged}. It should be noted that vulnerabilities with \emph{Scope} value “Changed” often score significantly higher. \cvss{3} splits exploitability into more precise dimensions, and the formula now reads as follows:
\begin{equation*}
    Exploitability_{v3} = 8.22\cdot av\cdot ac\cdot PR\cdot UI.
\end{equation*}
Note that \emph{Scope} does not appear directly, but changes the weight of \emph{PR}, which models the reality that privileges are more dangerous when crossing trust boundaries (\cite{CVSS_3.0}). The improved formula models phishing, client-side exploits, privilege escalation, and sandbox escapes, accurately. It also prevents exploitability from overwhelming impact and reduces the problem of "everything is a 10.0".
\cvss{3} divides the impact calculation based on the \emph{Scope} (\cite{CVSS_3.0}). In addition, all impact metrics get new possible values: "High", "Low", or "None". Then the formula for the impact sub-score (\emph{ISS}) is as follows:
\begin{equation*}
    ISS = 1-(1-C)(1-I)(1-A),
\end{equation*}
and the \emph{Scope}-aware formula for \emph{Impact} reads as follows
\begin{equation*}
Impact_{v3} =
\begin{cases}
6.42\cdot ISS, & \text{Scope unchanged}, \\[6pt]
7.52\cdot (ISS-0.029)-3.25\cdot (ISS-0.02)^{15},
& \text{Scope changed}. \\[6pt]
\end{cases}
\end{equation*}
This formula is a major improvement due to its non-linear penalty that prevents everything from scoring 10.0. It is also more accurate for modeling privilege escalation, VM/container escapes, microservice architectures, and cloud control plane compromise.
For the base score, there are three different formulas depending on \emph{Scope} and \emph{Impact} (\cite{CVSS_3.0}):
\begin{equation*}
Base_{v3} =
\begin{cases}
0, & \text{if Impact} \leq 0, \\[6pt]
\mathrm{round}\!\left(\min(\text{Impact} + \text{Exploitability}, 10)\right),
& \text{if Scope} = U, \\[6pt]
\mathrm{round}\!\left(\min\!\left(1.08 \cdot (\text{Impact} + \text{Exploitability}), 10\right)\right),
& \text{if Scope} = C.
\end{cases}
\end{equation*}
Compared to v2 formula, this one is better suited for cloud platforms, containers and VMs, privilege escalation chains, and client-side exploits. It produces a more realistic severity distribution and eliminates over-scoring driven by exploitability alone. It also forces analysts to think about trust boundaries.

The updated CVSS v3.1 was released in 2019 (\cite{CVSS_3.1}) keeping the scoring formulas intact but clarifying the metric definitions to reduce inconsistent scoring. The documentation and examples were improved and stronger guidance on how to score real-world vulnerabilities was provided. In addition, ambiguity was removed from the complexity of the attack, the required privileges, and the \emph{Scope} metrics. The v3.1 was created to reduce the scoring errors that were seen with v3.0, so that two people scoring the same vulnerability are more likely to get the same score. Currently, CVSS v3.1 is the recommended and most widely accepted version, although v3.0 is also acceptable but is becoming superseded. It is recommended to avoid v2 at least when making new assessments. A summary of key differences between CVSS versions is presented in Appendix~\ref{app:cvss_comparison}.

\im Threat Scores are a product of the \im service, a Lockheed Martin proprietary threat intelligence collection and analysis service that provides valuable threat intelligence to cyber analysts through a range of user interfaces. The \im service does this by maintaining a cyber data lake of publicly available sources, including threat intelligence reports, exploit code, indicators of active exploitation, official government advisories and other publications from CISA, FBI, NSA, and other foreign government entities just to name a few, to compute a daily Threat Score (1.0, 10.0) for every CVE in the NVD. The score correctly identifies 87.3\% of high threat CVEs listed in CISA's KEV catalog, which, as of August 2023, contained 984 entries ($\approx 0.5 \%$ of all CVEs) and focuses on Federal systems. Unlike the KEV catalog, \im is not limited by vendor acknowledgment, patch availability, or other constraints that can keep high risk CVEs out of the catalog.

Lockheed Martin stakeholders defined three requirement categories; practical, technical, and customer driven, to shape the scoring methodology. After evaluation, a Generalized Linear Model (GLM) was selected. In this model, the target variable is binary: whether a CVE appears in the KEV catalog. Predictor variables are cyber threat intelligence indicators that suggest active exploitation. Some benefits of the GLM equation includes being easy to measure, explainable with inputs and resulting scores, and able to easily extend the basic linear form to accommodate multiple predictors:

\begin{equation*}
y = m_1x_1 + m_2x_2 + \cdots + m_p x_p + b.
\end{equation*}

All inputs undergo feature engineering to become numeric.
From thousands of possible features, Pearson correlation testing narrowed the set to $\approx 70$ candidates, and iterative training refined it to 26 final features such as a CVE allowing escalation of privilege and an industry issued CVE exploitability warning (only four from the NVD). A score $\geq 7.0$ means the CVE's likelihood of active exploitation exceeds 70 \% of all CVEs. Scores in the 100th percentile (10.0) identify the most dangerous CVEs, which are highly probable to be in the KEV catalog. This approach delivers a consistent, data driven risk metric that extends beyond the limitations of existing federal listings. As new data sources are identified, reviewed and incorporated into the \im data lake, the scoring methodology and features used can be constantly adjusted to cover areas typically not included in KEV advisories like embedded systems.

EPSS (Exploit Prediction Scoring System) is an open data-driven scoring system designed to estimate the probability that a vulnerability could be exploited in the near future (\cite{EPSS}). It produces a score ranging from 0 to 1, where a higher score means that the vulnerability is more likely to be exploited. It focuses on the likelihood of exploiting the vulnerability, not on the impact or severity. It is updated daily based on new vulnerability data and observed exploit activity. Its intention is not to replace, but to complement, severity scores such as those of CVSS. The newest version (version 4) was released in 2025.

EPSS uses machine learning and historical exploit data to estimate the probability of a vulnerability being exploited (\cite{EPSS}). It gathers many types of data about each vulnerability, including CVE and metadata, references and contextual text, indicators of exploit code, and observed exploitation activity from sensors, IDS/IPS, honeypots, etc. It 
uses historical data to train a predictive model that learns which features correlate with the exploitation activity. The model is regularly retrained to incorporate recent trends and new exploitation patterns, and it scores every vulnerability to produce its probability of exploitation in the next 30 days. These scores are refreshed daily to adapt to new information, such as new exploit code, changes in exploit activity, or revised vulnerability information.

EPSS can be used to prioritize patching according to the likelihood of exploitation (\cite{EPSS}). For example, vulnerabilities with high EPSS scores may be patched earlier than those with lower EPSS scores, even if they share similar CVSS scores. Often, defenders combine CVSS for severity, EPSS for exploit likelihood, and other data to make balanced prioritization decisions. EPSS also supports analysis of effort, coverage, and efficiency. Different thresholds can be chosen based on risk tolerance and resources. Note that although CVSS and EPSS scores are related but fundamentally different as presented in Table \ref{tab:EPSS_vs_CVSS}. They can sometimes be correlated, but high severity does not guarantee high likelihood of exploiting and vice versa. 

By default, the EPSS score forecasts the probability of an exploitation in the next 30 days for each vulnerability.
Peripheral OT devices and software, such as those found in power substations, are unlikely targets for financially motivated threat actors, as exploitation efforts are difficult to monetize, which is reflected in very low EPSS scores for vulnerabilities in these assets. For a more informative comparison, we extended the forecast window to a full year by naively assuming the EPSS score to remain constant and accordingly extrapolated the 30-day estimate. This can be calculated via the complement:
\begin{equation*}
    P_{EPSS_{365}} = 1 - (1 - P_{EPSS_{30}}) ^ \frac{365}{30},
\end{equation*}
where $P_{EPSS_{30}}$ is the regular 30-day EPSS score.

\begin{table}[t]
\centering
\caption{Comparison between the EPSS and CVSS. }
\begin{tabular}{|l|l|l|}
\hline
\textbf{Feature} & \textbf{CVSS} & \textbf{EPSS} \\ \hline
What it scores & Vulnerability severity & Probability of real-world exploitation \\ \hline
Scale & 0-10 & 0-1 \\ \hline
Focus & Vulnerability characteristics & Observed and predicted threat likelihood \\ \hline
Based on & Defined metrics  & Data + ML model \\ \hline
Use & Risk/impact prioritization & Threat prioritization \\ \hline
\end{tabular}
\label{tab:EPSS_vs_CVSS}
\end{table}

Stakeholder-Specific Vulnerability Categorization (SSVC) is a framework that prioritizes decision making over scoring (\cite{SSVC}). It was developed in 2019 at Carnegie Mellon University (SEI) and is increasingly being used by governments and large enterprises. Its latest version (SSVC v2) is integrated into the Cybersecurity and Infrastructure Security Agency's (CISA) 2022 methodology (\cite{CISA}). SSVC asks about the likelihood of exploitation, severity, current threat activity, and system exposure, and maps this information to a decision (\cite{SSVC}). Thus, instead of giving a numerical score like CVSS, it considers a structured set of questions and outputs a clear action.

SSVC uses a decision tree model with a small set of factors (\cite{SSVC_v1}). Common factors include exploitation status (none, proof-of-concept exists, actively exploited), technical impact (partial, total compromise), exposure (system reachability), and mission/safety impact (effect on safety, operations, or critical services, and importance to organization). The exact factors vary slightly depending on the stakeholder, and different roles get slightly different decision trees. For example, software vendors focus on issuing patches and factors include customer impact and exploit maturity. System operators, including most enterprises, focus on patching systems, and the factors include exposure and business impact. Coordinators (e.g., CERTs) focus on publishing/alerting broadly. Because SSVC is intentionally designed to be stakeholder-specific and adaptable not to a single universal severity scale, the output categories vary depending on the users, e.g. CISAs model outputs one of 4 actions: Track (low priority), Track* (slightly higher concern), Attend (important) or Act (urgent) (\cite{CISA}).

The earlier version (v1) of SSVC is simpler and more conceptual in nature, such that the decision trees were understandable but sometimes ambiguous, and different teams could interpret nodes differently. SSVC v1 also mentioned different stakeholders, but they are not strongly separated. The biggest problem with v1 is that two analysts could evaluate the same vulnerability and get different results.

The second version (v2) is more refined, consistent, and scalable (\cite{SSVC_v2}). It is a formal decision model with each decision point having defined values and clear semantics. It is designed for automation and consistency to reduce “analyst interpretation drift”. Although both versions have similar input types, v2 refines them by better separating concerns, removing vague overlaps between factors, and cleaner definitions (e.g., what counts as “active exploitation”). The outputs in these two versions also look the same, but in v2 the definitions are more precise, easier to integrate into workflows, and mapped more cleanly to operational actions. Some implementations also refine timing expectations (e.g., immediate vs. scheduled) and Track / Track* distinctions. The biggest improvement with v2 is that it has more deterministic decision paths and is better for automation, policy enforcement, and auditing. In addition, v2 is designed to work more cleanly alongside CVSS and EPSS. The main differences between v1 and v2 are presented in Table \ref{tab:SSVC_versions}.
\begin{table}[tb]
\centering
\caption{Core differences between SSVC v1 and SSVC v2}
\label{tab:SSVC_versions}
\begin{tabular}{|l|l|p{.48\textwidth}|}
\hline
\textbf{Area} & \textbf{SSVC v1} & \textbf{SSVC v2} \\ \hline
Structure & Basic decision tree & Formalized decision model \\ \hline
Terminology & Less standardized & Clearly defined vocabulary \\ \hline
Stakeholder variants & Loosely defined & Explicitly separated (operator, supplier, coordinator) \\ \hline
Decision outputs & Track / Attend / Act & Same idea, but more rigorously defined \\ \hline
Factors & Fewer, less consistent & More precise and better scoped \\ \hline
Reproducibility & Moderate & Much higher \\ \hline
\end{tabular}
\label{tab:ssvc-v1-v2}
\end{table}

\subsection*{Sample Scenario: 2015 Ukraine Power Grid}

To evaluate and compare the scoring models described above, we applied them to the 2015 Ukraine Power Grid cyberattack, reconstructing the incident timeline using publicly available reports, primarily the E-ISAC/CIRT report (\cite{eisac_sans_ukraine_2016}). On 23 December 2015, the Ivano-Frankivsk region of Ukraine experienced a power outage lasting approximately three hours. Following the incident, malware was identified at multiple substations, including seven 110 kV and 23 × 35 kV substations. The outage affected up to 225,000 customers across three regional distribution companies and was attributed to one or more Advanced Persistent Threat (APT) groups. The attackers reportedly employed BlackEnergy and associated tools to compromise and simultaneously disrupt the targeted substations. This incident is particularly significant because it is widely regarded as the first confirmed power outage 
caused by a cyberattack. Based on the CIRT report, the major sequence of events (MSEL) include: (i)\emph{Phishing}, (ii)\emph{Network reconnaissance}, (iii)\emph{Active Directory compromise}, (iv)\emph{Establishment of an encrypted tunnel}, (v)\emph{SCADA HMI access}, (vi)\emph{Breaker manipulation}, (vii)\emph{Disruption of operator response}, and (viii)\emph{KillDisk malware deployment}.

\begin{figure}[tb]
        \centering
        \includegraphics[width=0.8\linewidth,trim=0 7mm 0 8mm,clip]{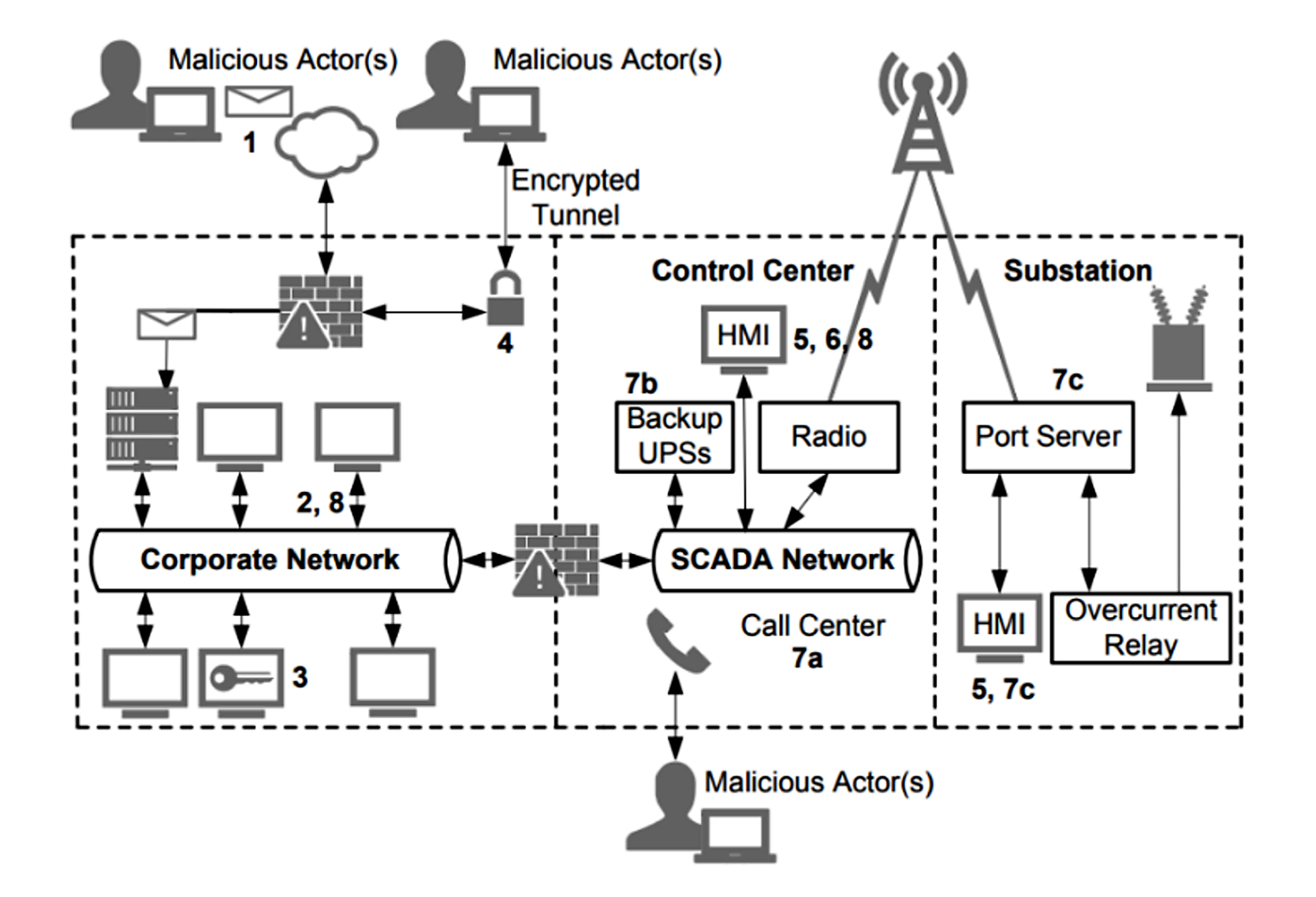}
        \caption{2015 Ukraine Power Grid Cyber Incident.}
        \label{fig:incident}
    \end{figure}

\begin{figure}[htpb]
    \centering
    \includegraphics[width=\textwidth,trim=1.0cm .7cm 1.9cm 0.2cm,clip]{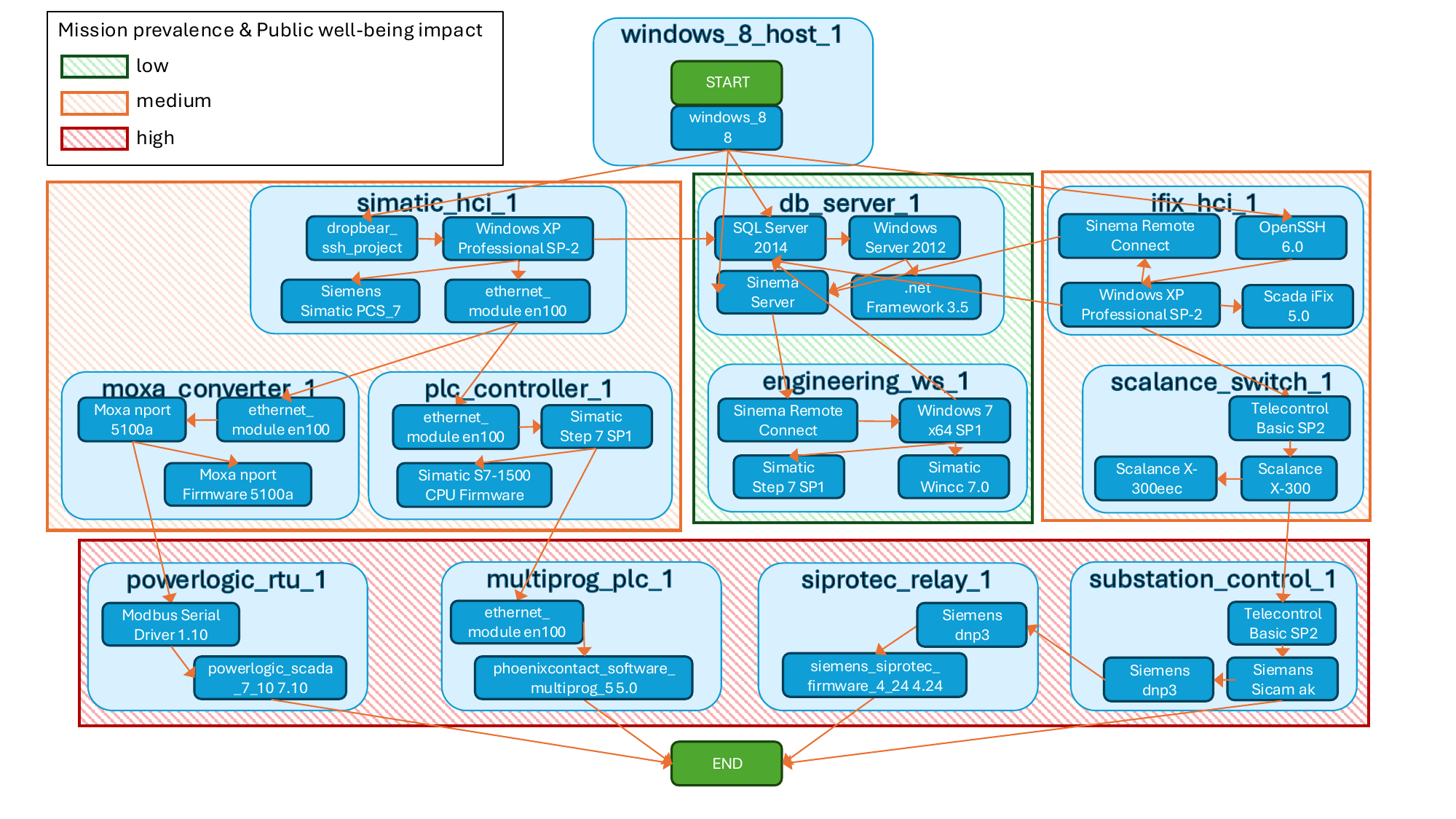}
    \caption{The assumed attack on Ukraine's energy infrastructure.}
    \label{fig:attack_graph}
\end{figure}

This sequence of events is mapped onto the representative power-grid architecture shown in Fig.~\ref{fig:incident}. Because the specific configuration of the Prykarpattya Oblenergo operational technology (OT) network was not disclosed in the CIRT report, we reconstructed a simplified topology based on the hardware and software assumptions described in the study (Fig.~\ref{fig:attack_graph}).

The left side of Fig.~\ref{fig:attack_graph} represents a configuration consistent with the technology available in 2015, including programmable logic controllers (PLCs) and host controller interfaces (HCIs). The right side represents a more contemporary configuration. This distinction enables us to assess whether multiple agents in \crow can identify attack vectors comparable to those documented in the actual incident under both legacy and modern system assumptions.

\subsection*{Methodologies}

To enable a numerical comparison between CVSS, EPSS, and SSVC outputs, a custom mapping layer is required to translate the categorical SSVC decisions into numerical values. Table~\ref{tab:ssvc-cvss-epss} presents the mapping approach adopted in this study. It should be emphasized that this mapping is heuristic in nature, as the other metrics capture fundamentally different aspects of vulnerability assessment. CVSS measures technical severity, EPSS estimates the probability of exploitation, and SSVC supports operational decision making through categorical actions. Consequently, EPSS values do not formally correspond to CVSS severity bands, and previous studies (\cite{Tan_2025}, \cite{Koscinski_2025}) have shown only a weak correlation between CVSS severity and EPSS exploitation probability. In practice, vulnerabilities with high CVSS scores may turn out to exhibit low EPSS values, while vulnerabilities with relatively low CVSS scores may still achieve high EPSS values when actively exploited in the wild. Therefore, the mapping presented in Table~\ref{tab:ssvc-cvss-epss} should be interpreted as an approximate operational alignment rather than a mathematically rigorous conversion between different metrics.

\begin{table}[t]
\centering
\caption{Operational comparison of SSVC decisions, CVSS ranges, and EPSS probabilities}
\begin{tabular}{|l|c|c|c|}
\hline
\textbf{SSVC Decision} & \textbf{Numeric Proxy} & \textbf{CVSS Range} & \textbf{EPSS Range} \\ \hline
Track   & 2 & 0.0 -- 3.9    & 0.00 -- 0.10 \\ \hline
Track*  & 5 & 4.0 -- 6.9    & 0.10 -- 0.30 \\ \hline
Attend  & 8 &  7.0 -- 8.9   & 0.30 -- 0.70 \\ \hline
Act     & 10 & 9.0 -- 10.0  & 0.70 -- 1.00 \\ \hline
\end{tabular}
\label{tab:ssvc-cvss-epss}
\end{table}

In order to obtain the SSVC estimates for each vulnerability in the network, we utilized the SSVC Coordinator decision tree, which requires determining the following decision points for each vulnerability: \emph{Exploitation status}, \emph{Technical impact}, \emph{Automatable}, and \emph{Mission prevalence \& Public well-being impact}.
\emph{Exploitation status}, representing whether a vulnerability is actively exploited, was derived using the VulnCheck Known Exploited Vulnerabilities (KEV) catalog.
\emph{Automatable} and \emph{Technical impact} values were derived from the CVSS metrics, whereas \emph{Mission prevalence \& Public well-being} was determined based on the node placement in the attack graph as shown in Fig.~\ref{fig:attack_graph}.
The full logic of how each of these values was derived is presented in Appendix~\ref{app:ssvc_logic}.

A common approach for combining vulnerability assessment metrics is the use of a weighted composite score. Instead of attempting to directly convert the outputs of different scoring models into a common scale, this method assigns each metric a weight corresponding to its relative importance and combines them into a single prioritization value. This approach acknowledges that vulnerability severity, exploitability, and remediation urgency represent distinct dimensions of risk and therefore contribute differently to decision making. In this study, a weighted composite score can be defined as
\begin{equation*}
    \text{Risk Score} = w_1 * \text{CVSS} + w_2 * \text{EPSS} + w_3 * \text{SSVC},
\end{equation*}
where $w_1$, $w_2$, and $w_3$ are weighting coefficients and $w_1 + w_2 + w_3 = 1$. The CVSS component represents the technical severity of a vulnerability, EPSS provides an estimate of the likelihood of exploitation, and SSVC contributes an operational perspective by incorporating contextual factors related to remediation decisions. Since SSVC produces categorical outputs, a numerical mapping must first be applied to enable integration into the composite score. The primary advantage of a weighted composite score is its flexibility. Organizations can adjust the weights to reflect their risk management objectives and operational environment. For example, environments primarily concerned with immediate threats may assign a higher weight to EPSS, while organizations operating critical infrastructure may place greater emphasis on SSVC-based decision outcomes. Similarly, a higher CVSS weight may be appropriate when technical impact is considered the dominant factor. However, the selection of weighting coefficients introduces a degree of subjectivity, and different weight combinations can lead to substantially different prioritization results. Furthermore, the approach assumes that the three metrics can be meaningfully combined despite measuring different characteristics of a vulnerability. Thus, a weighted composite score should be viewed as a prioritization aid rather than an objective measure of risk.

An alternative approach is to use CVSS as the baseline severity measure and adjust it using factors derived from SSVC and EPSS to derive a priority score:

\begin{equation*}
    \text{Priority Score} = \text{CVSS} * \frac{\text{SSVC}}{10} * \text{EPSS modifier},
\end{equation*}
where CVSS represents the technical severity of the vulnerability, SSVC is a numerical coefficient derived from the SSVC decision category, EPSS modifier = 1 + EPSS, and EPSS is the estimated probability of exploitation.  Unlike the weighted composite score, this method preserves the interpretation of CVSS as the primary severity metric while incorporating operational urgency and exploitation likelihood as multiplicative adjustments. As a result, vulnerabilities with high severity but low operational urgency receive a lower priority score, whereas vulnerabilities with active exploitation and urgent remediation requirements receive a higher score. This behavior aligns more closely with practical vulnerability management processes, where remediation decisions are influenced not only by technical severity but also by the current threat landscape and organizational context. One advantage of this approach is its simplicity and interpretability. The contribution of each component can be easily understood, and the resulting score naturally reflects the interaction between severity, exploitability, and remediation urgency. In particular, the multiplicative formulation emphasizes situations in which multiple risk factors are simultaneously high, thereby highlighting vulnerabilities that are both severe and likely to be exploited. However, the method also has limitations. The selection of SSVC coefficients remains subjective and may vary between organizations. Furthermore, the formula assumes that the effects of severity, exploitability, and decision urgency interact multiplicatively, which may not accurately represent real-world risk. Consequently, the resulting priority score should be regarded as a prioritization heuristic rather than a formally validated measure of vulnerability risk.

\subsection*{Dataset Description}

The dataset consists of 659 vulnerability records collected from 12 information systems and 27 software products. Each record represents a software vulnerability identified by a Common Vulnerabilities and Exposures (CVE) identifier and includes multiple vulnerability assessment metrics derived from established cybersecurity scoring frameworks. The dataset contains 592 unique CVEs, indicating that some vulnerabilities affect multiple systems or software components.

For each vulnerability, the dataset includes scores from the CVSS versions 2.0 and 3.0. Specifically, the dataset contains the \cvss{2} Base Score, Exploitability Subscore, and Impact Subscore, as well as the corresponding \cvss{3} metrics. These metrics provide standardized assessments of vulnerability severity, exploitability, and potential impact.

In addition to CVSS-based measures, the dataset incorporates EPSS metrics, which 
represents the probability that a vulnerability will be exploited in the wild, while \yepss captures a longer-term exploitation likelihood estimate over a period of 365-days. These metrics complement severity-based assessments by introducing a predictive perspective on real-world exploitation risk.

The dataset also includes SSVC recommendations, the labels of which indicate recommended response actions, including Track, Track*, Attend, and Act. The majority of records are classified as Track (621 instances), while smaller numbers are categorized as Attend (27 instances), Track* (9 instances), and Act (2 instances).

Overall, the dataset combines severity-oriented (CVSS), likelihood-oriented (EPSS), and decision-support (SSVC) vulnerability assessment frameworks, enabling comparative analyses of different scoring systems and supporting research on vulnerability prioritization and risk-based remediation strategies.

\section{Results}

A total of 113 vulnerabilities with complete \cvss{2} and \cvss{3} scoring information were analyzed the statistics of which are shown in Table \ref{tab:statistics_sub} and reveal 
some notable differences between the two scoring systems. The mean \cvss{2} Base Score was 6.37 ($SD=2.25$), while for the mean \cvss{3} it was 7.55 ($SD = 1.50$), indicating that \cvss{3} generally assigns higher overall severity scores. Statistics of all vulnerabilities for which vulnerability scoring were found can be found in Appendix \ref{app:stats_all}. 

\begin{table}[t]
\centering
\caption{Statistics of the 113 vulnerabilities for which both \cvss{2} and \cvss{3} scores were found.}
\label{tab:statistics_sub}
\begin{tabular}{lcccccccc}
\hline
 & count & mean & std & min & 25\% & 50\% & 75\% & max \\
\hline
\cvss{2} Base Score & 113 & 6.37 & 2.25 & 2.10 & 4.40 & 5.80 & 9.00 & 10.00 \\
\cvss{2} Exploitability & 113 & 8.26 & 2.02 & 3.40 & 8.00 & 8.60 & 10.00 & 10.00 \\
\cvss{2} Impact & 113 & 6.03 & 3.01 & 2.90 & 2.90 & 6.40 & 10.00 & 10.00 \\
\cvss{3} Base Score & 113 & 7.55 & 1.50 & 3.70 & 6.50 & 7.50 & 8.80 & 9.80 \\
\cvss{3} Exploitability & 113 & 2.89 & 0.92 & 0.80 & 2.20 & 2.80 & 3.90 & 3.90 \\
\cvss{3} Impact & 113 & 4.58 & 1.56 & 1.40 & 3.60 & 5.90 & 5.90 & 6.00 \\
EPSS & 113 & 0.17 & 0.27 & 0.00 & 0.01 & 0.03 & 0.17 & 0.99 \\
\yepss & 113 & 0.47 & 0.37 & 0.02 & 0.13 & 0.35 & 0.90 & 1.00 \\
\im & 113 & 7.55 & 2.11 & 1.80 & 6.30 & 7.90 & 8.70 & 10.00 \\
SSVC$_{\text{NUM}}$ & 113 & 2.82 & 2.01 & 2.00 & 2.00 & 2.00 & 2.00 & 8.00 \\
\hline
\end{tabular}
\end{table}

The largest differences were observed in the Exploitability metric as \cvss{2} produced a mean Exploitability Score of 8.26, compared with 2.89 for \cvss{3}. The mean absolute difference between the exploitability scores was 5.37 points (SD = 1.42), suggesting substantial methodological differences in the way that 
exploitability is assessed.

Impact scores were more closely aligned, although meaningful differences remained. The mean Impact Score decreased from 6.03 in \cvss{2} to 4.58 in \cvss{3} and the average absolute difference was 1.92 points (SD = 1.60).

For the overall Base Score, the mean absolute difference between versions was 1.52 points (SD = 1.19), with a maximum observed difference of 5.70 points. These results demonstrate that the two scoring systems frequently produce different severity assessments for the same vulnerability.

EPSS assigns very low exploitation probabilities to most of the vulnerabilities. The median EPSS score is only 0.03, and the first quartile is 0.01, indicating that at least half of the vulnerabilities have relatively low predicted probability of exploitation. However, the distribution is highly right-skewed: the mean of 0.17 is substantially higher than the median, and the maximum reaches 0.99. Thus, while most vulnerabilities have low predicted exploitation probability, a small subset receives very high EPSS values. This distinction is important for prioritization because an average EPSS value alone would obscure the existence of vulnerabilities with substantially elevated exploitation likelihood.

Extending the EPSS measure to a 365-day horizon produces considerably higher values. The median increases from 0.03 to 0.35, while the mean rises from 0.17 to 0.47. The 75th percentile similarly increases from 0.17 to 0.90. This indicates that the apparent risk represented by EPSS is strongly dependent on the forecasting horizon. A vulnerability that appears unlikely to be exploited in the short term may nevertheless have a substantial cumulative likelihood of exploitation over a longer period. Therefore, EPSS should be interpreted as a temporal prediction of exploitation likelihood rather than as a direct measure of vulnerability severity.

\im produces a markedly different picture. Its mean score of 7.55 is identical to the mean \cvss{3} base score, but the distributions are fundamentally different. \im has a higher median (7.90), with 75\% of observations at or below 8.70 with a substantially larger standard deviation (2.11). Its scores range from 1.80 to 10.00, which indicates that \im differentiates more strongly between vulnerabilities than \cvss{3} in this dataset. Thus, the similarity between the two means is potentially misleading: the same mean value can arise from very different distributions and does not imply that the two frameworks rank vulnerabilities similarly.

SSVC displays the strongest concentration of any metric in the table. Its median, first quartile, and third quartile are all 2.00, while the mean is 2.82 and the maximum is 8.00. This means that at least 75\% of the observations receive the same numerical value of 2, with a relatively small number of higher-valued observations pulling the mean upward. The resulting distribution suggests that SSVC provides considerably less differentiation among the vulnerabilities in this particular dataset than the continuous scoring systems.

Overall, the table demonstrates that the scoring frameworks do not provide interchangeable representations of vulnerability risk. \cvss{2} and \cvss{3} produce different severity distributions despite being successive versions of the same standard, while EPSS focuses on predicted exploitation likelihood, \im incorporates threat-oriented information, and SSVC reduces vulnerabilities into a smaller set of decision-oriented categories. The particularly large difference between \cvss{2} and \cvss{3} exploitability scores illustrates why comparing individual component scores across versions can be misleading. More importantly, the different distributions of EPSS, \im, and SSVC indicate that disagreement between frameworks is not necessarily evidence that one framework is incorrect; rather, the frameworks capture different dimensions of risk.

\begin{figure}[htb]
    \centering
    \includegraphics[width=.7\textwidth,trim=0 5mm 0 4mm,clip]{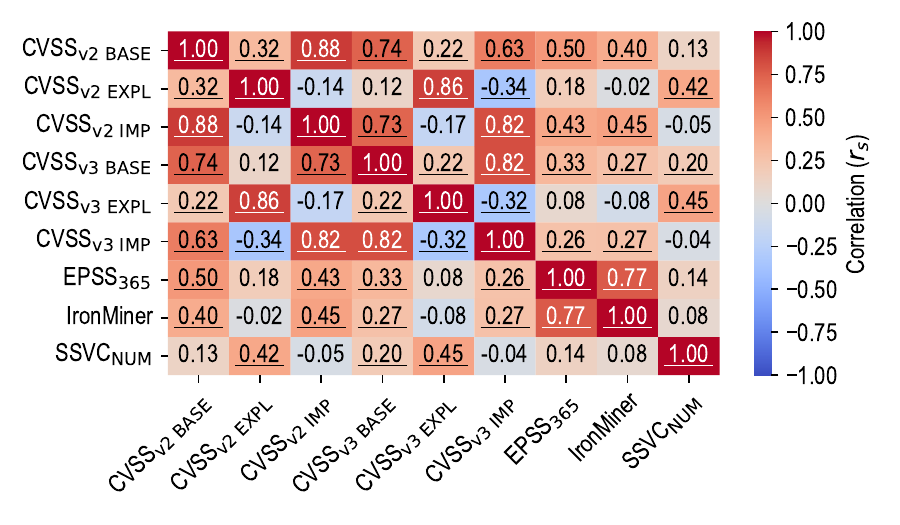}
    \caption{Spearman correlation between scoring systems. The \ul{underlining} indicates \mbox{$r_s$-values} for which Benjamini-Hochberg corrected $p$-value is $<0.05$.}
    \label{fig:spearman}
\end{figure}

Figure \ref{fig:spearman} presents the Spearman rank correlation matrix among the \cvss{2} and \cvss{3} Base, Exploitability, and Impact scores, as well as \yepss, \im, and SSVC. The correlation analysis demonstrates strong agreement between \cvss{2} and \cvss{3}, particularly for corresponding Impact ($r_s = 0.82$) and Exploitability ($ r_s = 0.86$) scores. \yepss and \im are also strongly correlated ($r_s = 0.77$), but exhibit only moderate associations with CVSS, suggesting that they capture complementary information related to vulnerability prioritization. In contrast, SSVC shows generally weak correlations with the other measures, indicating a substantially different prioritization perspective. Overall, the results suggest that combining severity-based CVSS measures with exploitation- and decision-oriented approaches such as EPSS, \im, and SSVC may provide a more diverse assessment of vulnerability priority.

\section{Conclusions}

This study examined the behavior of four vulnerability scoring systems (CVSS, EPSS, SSVS, and \im) using a reconstructed topology of the 2015 Ukraine Power Grid cyber incident. The results demonstrate that these systems should not be treated as interchangeable measures of vulnerability risk. Even successive versions of the same framework, \cvss{2} and \cvss{3}, produced substantially different score distributions, particularly for exploitability. More broadly, EPSS and \im primarily capture exploitation likelihood, CVSS emphasizes technical severity and impact, while SSVC provides a decision-oriented assessment incorporating operational context.

The correlation analysis further shows that the differences between scoring systems are structured rather than simply random. \cvss{2} and \cvss{3} exhibit strong correlations across corresponding measures, particularly for exploitability and impact, indicating substantial consistency between the two CVSS versions despite differences in their scoring methodologies. \yepss and \im also show a strong correlation, suggesting that these approaches capture related exploitation-oriented characteristics. However, their correlations with CVSS are generally moderate, indicating that exploitation likelihood is not fully represented by technical severity scores. SSVC exhibits comparatively weak correlations with both CVSS and the exploitation-oriented measures, particularly \yepss and \im, reinforcing its distinct, decision-oriented perspective. Overall, the correlation structure indicates that the scoring systems capture overlapping but non-equivalent dimensions of vulnerability risk, supporting the use of multiple assessment signals rather than treating any individual score as a complete representation of vulnerability priority.

Our study supports the broader use of multi-signal vulnerability assessment in operational environments. Combining severity, likelihood of exploitation, and operational decision information has the potential to provide a more informative basis for remediation decisions than any individual scoring system. The weighted composite and CVSS-anchored priority formulations presented in this work provide possible mechanisms for such integration; however, their weighting, mapping, and interaction assumptions remain heuristic and should not yet be interpreted as formally validated measures of risk. A more detailed analysis of the proposed Risk Score and Priority Score is left for future work. Such work should evaluate how different parameterizations affect vulnerability rankings and defensive prioritization, and should assess the robustness of these formulations across additional OT/ICS environments and incident topologies. Future research can therefore build on the present finding that vulnerability scoring systems capture distinct and complementary signals to determine how those signals can be combined most effectively for operational vulnerability prioritization.

\bibliographystyle{splncs04}
\bibliography{references}

\begin{thebibliography}{10}
\providecommand{\url}[1]{\texttt{#1}}
\providecommand{\urlprefix}{URL }
\providecommand{\doi}[1]{https://doi.org/#1}

\bibitem{SSVC}
Carnegie Mellon University, \url{https://certcc.github.io/SSVC/}: SSVC: Stakeholder-Specific Vulnerability Categorization (nd)

\bibitem{CISA}
Cybersecurity and Infrastructure Security Agency: CISA Stakeholder-Specific Vulnerability Categorization Guide (November 2022)

\bibitem{eisac_sans_ukraine_2016}
{E-ISAC}, {SANS ICS}: Analysis of the cyber attack on the {Ukrainian} power grid: Defense use case. Tech. rep., Electricity Information Sharing and Analysis Center (E-ISAC) and SANS Industrial Control Systems (March 2016)

\bibitem{CVSS_3.0}
FIRST, \url{https://www.first.org/cvss/v3.0/specification-document}: Common Vulnerability Scoring System v3.0: Specification Document (nd)

\bibitem{CVSS_3.1}
FIRST, \url{https://www.first.org/cvss/v3.1/specification-document}: Common Vulnerability Scoring System v3.1: Specification Document (nd)

\bibitem{EPSS}
FIRST, \url{https://www.first.org/epss/}: Exploit Prediction Scoring System (EPSS) (nd)

\bibitem{Koscinski_2025}
Koscinski, V., Nelson, M., Okutan, A., Falso, R., Mirakhorli, M.: Conflicting scores, confusing signals: An empirical study of vulnerability scoring systems. In: Proceedings of the 2025 ACM SIGSAC Conference on Computer and Communications Security. p. 1904–1918. CCS '25, Association for Computing Machinery, New York, NY, USA (2025). \doi{10.1145/3719027.3765210}, \url{https://doi.org/10.1145/3719027.3765210}

\bibitem{CVSS_2.0}
Mell, P., Romanosky, S., Scarfone, K.: A Complete Guide to the Common Vulnerability Scoring System Version 2.0. \url{https://www.first.org/cvss/v2/guide} (2007)

\bibitem{Pinaev26}
Pinaev, A.: Models and methods for prioritizing software vulnerabilities based on business-criticality indicators and probability of exploitation. International Journal of Modern Computer Science and IT Innovations  \textbf{3}(4),  7--16 (2026). \doi{https://doi.org/10.55640/ijmcsit-v03i04-01}

\bibitem{SSVC_v2}
Spring, J., Hatleback, E., Householder, A., Manion, A., Oliver, M., Sarvapalli, V., Tyzenhaus, L., Yarbrough, C.: Prioritizing vulnerability response: A stakeholder-specific vulnerability categorization (version 2.0) (April 2021)

\bibitem{SSVC_v1}
Spring, J., Hatleback, E., Householder, A., Manion, A., Shick, D.: Prioritizing vulnerability response: A stakeholder-specific vulnerability categorization (version 1.1) (December 2020)

\bibitem{Tan_2025}
Tan, E.H., Unal, I.E., Rhea, S., Tatar, U.: Analysis of vulnerability severity and exploit probability scoring frameworks: Cvss and epss. In: 2025 Systems and Information Engineering Design Symposium (SIEDS). pp. 54--59 (2025). \doi{10.1109/SIEDS65500.2025.11021216}

\bibitem{Wunder24}
Wunder, J., Kurtz, A., Eichenmüller, C., Gassmann, F., Benenson, Z.: Shedding light on cvss scoring inconsistencies: A user-centric study on evaluating widespread security vulnerabilities. In: 2024 IEEE Symposium on Security and Privacy (SP) (2024). \doi{10.1109/SP54263.2024.00058}

\end{thebibliography}

\appendix
\clearpage
\section{Summary of differences between CVSS 2.0 and 3.x}\label{app:cvss_comparison}
\begin{table}[h]
\centering
\caption{Comparison between CVSS 2.0 and CVSS 3.x across key aspects.}
\begin{tabular}{|l|l|l|}
\hline
\textbf{Aspect} & \textbf{CVSS 2.0} & \textbf{CVSS 3.x} \\
\hline

\multicolumn{3}{|l|}{\textbf{Exploitability}} \\

\hline
Formula Constant   & 20              & 8.22                    \\
Number of Metrics  & 3               & 4 (+ scope influence)   \\
Authentication     & Explicit metric & Removed                 \\
Privilege Modeling & Indirect        & Explicit                \\
User Interaction   & Not modeled     & Explicit                \\
Scope Awareness    & None            & Indirect (PR weighting) \\
Realism            & Lower           & Much higher             \\
\hline

\multicolumn{3}{|l|}{\textbf{Impact}} \\

\hline
Scope Awareness       & \color{red}{\tikzxmark} & \color{green}\tikzcmark                              \\
Non-Linear Behavior   & \color{red}{\tikzxmark} & \color{green}\tikzcmark \color{black}(scope changed) \\
Cross-Boundary Impact & \color{red}{\tikzxmark} & Explicit                                             \\
Max Impact Scaling    & 10.41                   & 6.42/variable                                        \\
\hline

\multicolumn{3}{|l|}{\textbf{Base Score}} \\

\hline
Combination method    & Weighted sum            & Direct sum              \\
Impact Weight         & 60\%                    & Implicitly dominant     \\
Exploitability Weight & 40\%                    & Additive, capped        \\
Scope Awareness       & \color{red}{\tikzxmark} & \color{green}\tikzcmark \\
Non-Linear Components & \color{red}{\tikzxmark} & \color{green}\tikzcmark \\
Scaling               & Fixed constants         & Conditional (1.08)      \\
Cap at 10             & Implicit                & Explicit                \\
\hline

\multicolumn{3}{|l|}{\textbf{General}} \\

\hline
Release Year               & 2007                    & 2015 (3.0), 2019 (3.1)                      \\
Score Range                & 0.0-10.0                & 0.0-10.0                                    \\
Scope Metric               & \color{red}{\tikzxmark} & \color{green}\tikzcmark                     \\
User Interaction Metric    & \color{red}{\tikzxmark} & \color{green}\tikzcmark                     \\
Privileges Required Metric & \color{red}{\tikzxmark} & \color{green}\tikzcmark                     \\
Physical Attacks           & Limited                 & Explicit                                    \\
Formula Changes            & -                       & Major                                       \\
Clarity \& Consistency     & Low                     & Medium (3.0), High (3.1)                    \\
Current Recommendation     & \color{red}{\tikzxmark} & {\color{green}\tikzcmark} (3.1 recommended) \\
\hline
\end{tabular}
\end{table}

\clearpage
\section{SSVC decision point derivation logic}\label{app:ssvc_logic}
\begin{table}[h]
\centering
\caption{SSVC decision point derivation logic.}
\label{tab:ssvc_derivation}
\footnotesize
\begin{tabular}{@{} >{\raggedright\arraybackslash}p{\colA} >{\raggedright\arraybackslash}p{\colB} >{\raggedright\arraybackslash}p{\colC} >{\raggedright\arraybackslash}p{\colD} >{\raggedright\arraybackslash}p{\colE} @{}}
\hline
\textbf{Decision Point} & \textbf{Value} & \multicolumn{3}{l}{\textbf{Criteria}} \\
\hline

\multirow{8}{*}{\shortstack[l]{Exploitation\\Status}}
  & \multirow{4}{*}{\emph{active}}
    & \WideCellThree{Listed in CISA KEV} \\
  & & \WideCellThree{Observed in \emph{VulnCheck Canary honeypots}} \\
  & & \WideCellThree{Utilized in a ransomware campaign} \\
  & & \WideCellThree{Otherwise reported exploited by VulnCheck} \\
\cline{2-5}
  & \multirow{2}{*}{\emph{poc}}
    & \WideCellThree{Exploit in VulnCheck exploit database} \\
  & & \WideCellThree{Description contains keywords, such as \emph{metasploit} or \emph{proof of concept}} \\
\cline{2-5}
  & \emph{none} & \WideCellThree{Otherwise} \\
\hline

\multirow{5}{*}{Automatable}
  & & \textit{CVSSv2} & \textit{CVSSv3} & \textit{Value} \\
\cline{3-5}
  & \multirow{3}{*}{\emph{yes} (all of)}
    & \texttt{accessVector}     & \texttt{attackVector}     & \texttt{NETWORK} \\
  & & \texttt{accessComplexity} & \texttt{attackComplexity} & \texttt{LOW} \\
  & & \texttt{authentication}   & \texttt{userInteraction}  & \texttt{NONE} \\
\cline{2-5}
  & \emph{no} & \WideCellThree{Otherwise} \\
\hline

\multirow{5}{*}{\shortstack[l]{Technical\\Impact}}
  & & \textit{Metric} & \textit{CVSSv2} & \textit{CVSSv3} \\
\cline{3-5}
  & \multirow{3}{*}{\emph{total} (any of)}
    & \texttt{confidentialityImpact} & \texttt{COMPLETE} & \texttt{HIGH} \\
  & & \texttt{integrityImpact}       & \texttt{COMPLETE} & \texttt{HIGH} \\
  & & \texttt{availabilityImpact}    & \texttt{COMPLETE} & \texttt{HIGH} \\
\cline{2-5}
  & \emph{partial} & \WideCellThree{Otherwise} \\
\hline

\multirow{5}{*}{\shortstack[l]{Mission\\prevalence \\\&\\Public\\well-being}}
  & \multirow{5}{*}{\shortstack[l]{low, medium,\\or high}}
    & \multirow{5}{*}{\parbox{\dimexpr\colC+\colD+\colE+4\tabcolsep\relax}{Based on node placement in the attack graph (see Fig.~\ref{fig:attack_graph})}} \\
  & & \\
  & & \\
  & & \\
  & & \\
\hline
\end{tabular}
\end{table}

\clearpage
\section{Statistics of all the vulnerabilities}\label{app:stats_all}
\begin{table}[h!]
\centering
\caption{Statistics of all vulnerabilities for which vulnerability scoring were found.}
\begin{tabular}{lcccccccc}
\hline
Column & count & mean & std & min & 25\% & 50\% & 75\% & max \\
\hline
\cvss{2} Base Score & 647 & 6.79 & 2.14 & 1.20 & 5.00 & 7.20 & 9.30 & 10.00 \\
\cvss{2} Exploitability & 647 & 7.07 & 2.60 & 1.90 & 3.90 & 8.60 & 8.60 & 10.00 \\
\cvss{2} Impact & 647 & 7.41 & 2.89 & 2.90 & 4.90 & 6.90 & 10.00 & 10.00 \\
\cvss{3} BaseScore & 125 & 7.51 & 1.53 & 3.70 & 6.50 & 7.50 & 8.80 & 9.80 \\
\cvss{3} Exploitability & 125 & 2.86 & 0.93 & 0.80 & 2.20 & 2.80 & 3.90 & 3.90 \\
\cvss{3} Impact & 125 & 4.56 & 1.55 & 1.40 & 3.60 & 5.90 & 5.90 & 6.00 \\
EPSS & 659 & 0.15 & 0.22 & 0.00 & 0.02 & 0.04 & 0.20 & 1.00 \\
\yepss & 659 & 0.52 & 0.35 & 0.02 & 0.19 & 0.40 & 0.93 & 1.00 \\
\im & 659 & 6.44 & 1.99 & 1.50 & 5.00 & 6.30 & 7.80 & 10.00 \\
SSVC$_{\text{NUM}}$ & 659 & 2.31 & 1.30 & 2.00 & 2.00 & 2.00 & 2.00 & 10.00 \\
\hline
\end{tabular}
\end{table}

\end{document}